\documentclass[pra, twocolumn , showpacs
			     ,nofootinbib 
				         	]{revtex4}
\usepackage{amsmath}
\usepackage{amssymb}
\usepackage{bm}
\usepackage{graphicx}
\usepackage{subfigure}
\usepackage{dcolumn}

\newcommand{\bo}{\hat{b}^{\phantom\dag}}
\newcommand{\ba}{\hat{b}^{\dag}}
\newcommand{\fo}{\hat{f}^{\phantom\dag}}
\newcommand{\fa}{\hat{f}^{\dag}}
\newcommand{\co}{\hat{c}^{\phantom\dag}}
\newcommand{\ca}{\hat{c}^{\dag}}

\newcommand{\no}{\hat{n}}
\newcommand{\Ho}{\hat{H}}

\newcommand{\Po}{\hat{P}}

\newcommand{\So}{\hat{\bm S}}

\newcommand{\la}{\langle}
\newcommand{\ra}{\rangle}
\newcommand{\be}{\begin{equation}}
\newcommand{\ee}{\end{equation}}
\newcommand{\bes}{\begin{eqnarray}}
\newcommand{\ees}{\end{eqnarray}}

\newcommand{\br}{{\bm r}}

\newcommand{\bk}{{\bm k}}

\newcommand{\dw}{\downarrow}
\newcommand{\up}{\uparrow}

\begin{document}

\title{Controlled hole doping of a Mott insulator of ultracold fermionic atoms}
\author{Andr\'e~Eckardt$^1$}
\email{andre.eckardt@icfo.es}
\author{Maciej~Lewenstein$^{1,2}$}
\affiliation{$^1$ICFO-Institut de Ci\`encies Fot\`oniques, 
Av.\ Canal Ol\'impic s/n, E-08860 Castelldefels (Barcelona), Spain,}
\affiliation{$^2$ICREA-Institucio Catalana de Recerca i Estudis Avan\c{c}ats,
Lluis Companys 23, E-08010 Barcelona, Spain}
\date{July 14, 2010}

\begin{abstract}
Considering a system of ultracold atoms in an optical lattice, we propose a
simple and robust implementation of a quantum simulator for the 
\emph{homogeneous} $t$-$J$ model with a well-controlled fraction of holes $x$.
The proposed experiment can provide valuable insight into the physics of cuprate
superconductors. 
A similar scheme applied to bosons, moreover, allows one to investigate
experimentally the subtle role of inhomogeneity when a system passes from one
quantum phase to another.
\end{abstract}

\maketitle
Gases of ultracold atoms in optical lattice potentials provide extremely clean
and controllable conditions for studying strongly correlated many-body physics
\cite{Lattices}. Since for deep lattices these systems are described
quantitatively by Hubbard-type Hamiltonians \cite{JakschEtAl98}, they have great
potential to serve as quantum simulators \cite{Feynman82} for
paradigmatic models of condensed matter physics. This prospect is fed by
results for bosonic systems. In a seminal experiment, the quantum phase
transition from a superfluid of bosons to a strongly correlated Mott
insulator---predicted for the bosonic Hubbard model \cite{FisherEtAl89}---has
been observed \cite{GreinerEtAl02}. Moreover, quantitative
agreement between experiment and \emph{ab initio} quantum Monte Carlo
simulations clearly confirm the validity of the Bose-Hubbard description
\cite{TrotzkyEtAl09}. 

While the elementary Hubbard model for bosons is rather well understood,
this is not the case for repulsively interacting fermions: In a Mott insulator,
with interaction localizing one particle of ``spin'' $s=$~$\up$~or~$\dw$ at each
site, Fermi statistics gives rise to an \emph{antiferromagnetic} superexchange
coupling $J$ between neighboring spins. Intriguing physics is expected when such
a quantum antiferromagnet is frustrated, either by a non-bipartite lattice
geometry \cite{Antiferro} or by doping it with holes (or fermions) displacing
spins when moving around \cite{LeeEtAl06}.
The latter scenario for a square lattice is conjectured to give rise to
$d_{x^2-y^2}$-wave pair superfluidity and to explain basic properties of
high-temperature cuprate superconductors \cite{Anderson87,AndersonEtAl04,
LeeEtAl06}. However, conclusive theoretical evidence of whether the plain
Hubbard physics supports a superconducting state is still lacking. As
pointed out already in Ref.~\cite{HofstetterEtAl02}, here a cold atom
realization of the fermionic Hubbard model (or its descendant, the $t$-$J$
model) could provide critical insight.

The recent observation of a Mott insulator of repulsively interacting fermionic
atoms in a deep optical lattice \cite{Mott} is an important first step toward a
clean cold atom realization of strongly correlated fermionic Hubbard physics.
However, further steps in that direction require solutions to two problems: 
(i) The temperatures that can be achieved presently are still larger than (or at
most comparable to) the energy scale of the superexchange spin-spin coupling
\cite{AdiabaticCooling}. Promising novel cooling procedures have been
proposed to tackle this problem \cite{AdvancedCooling} (cf.\ also 
\cite{EckardtEtAl09b}).
(ii)  
While it is relatively easy to create an incompressible Mott region with one
atom per site in the center of a parabolic trap, it is very hard to dope such
a trapped Mott insulator in a controlled way. When the particle number
in the center of the trap is lowered, e.g.\ by slowly widening the trap, the
central Mott-insulator phase will not be doped with holes homogeneously. Rather,
it will melt from the edge. Also simply switching off the trapping potential 
(i.e.\ compensating it with a blue-detuned laser) is difficult, since then
particles may leak out and it will be hard to control the lattice filling.
In this Rapid Communication we propose a robust solution to problem (ii). 
Our method
allows one to both accurately control the fraction $x$ of hole doping and 
effectively compensate the trapping potential. Both are crucial in order to learn
about high-temperature superconductivity with ultracold atoms. Applied to bosons,
our scheme, moreover, allows one to experimentally investigate the non-trivial
influence of spatial inhomogeneity when a system passes from one quantum phase
to another \cite{Trap}.

\begin{figure}[t]\centering
\includegraphics[width = 0.8\linewidth]{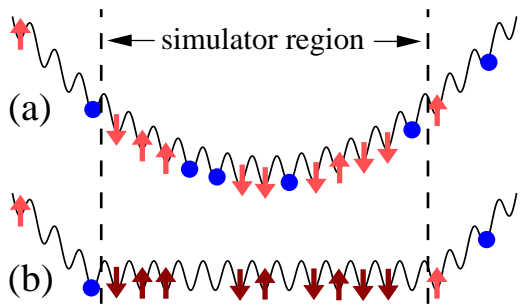}
\caption{\label{fig:principle} (color online) (a) Sketch of the lattice and
trapping potential felt by both fermions and bosons. Assuming strong on-site
repulsion and symbolizing fermions by light red arrows and bosons by blue 
bullets, an occupation snap shot is sketched. There is a central mixed
Mott region, the simulator region, with one boson or fermion per site. 
(b) In the simulator region, fermions move by fermion-boson swaps not changing
potential energy. Here a description solely in terms of a new type of
fermions (dark red arrows) is possible, with bosons implicitly taken care of as
holes. These fermions move with an effective tunneling amplitude and do not
feel the trap; the fraction $x$ of holes equals that of bosons.}
\end{figure}

Our basic idea is to create holes in a fermionic Mott insulator by adding
auxiliary bosonic particles to the system that interact repulsively
both with the fermions and with each other. Each boson repels a fermion from one
site. Experimentally, one has to create a mixed Mott phase with one particle,
either fermion or boson, per site. 
This spatially rather well-defined ``simulator region'' resembles the system to
be simulated. The degree of hole
doping is well controlled by the number ratio between bosons and fermions. 
The price to be payed is that, unlike the tunneling of a real hole, the motion
of the bosons happens via boson-fermion swaps, being slow second-order
superexchange processes (cf.~\cite{KuklovSvistunov03,LewensteinEtAl04} for the
case of ``spinless'' Bose-Fermi mixtures). But this type of hole kinetics
brings also a great advantage: Consider the trap being very similar for fermions
and bosons (think of two Ytterbium isotopes in a far off-resonant dipole
potential). Then such a boson-fermion swap on two neighboring sites will change
the potential energy only marginally, even if the trapping potential does change
between both sites. Hence, within the simulator region the physics is hardly
influenced by the trap, cf.~Fig.~\ref{fig:principle}. Below we
will show that this region is accurately described by the \emph{homogeneous}
$t$-$J$ model with \emph{controlled hole fraction}~$x$. The $t$-$J$ model
\cite{ZouAnderson88} (cf.~\cite{Duan05Anna} for other cold atom applications)
describes the doped fermionic Mott-insulator, with fermion hopping $t$ and
superexchange spin coupling $J$ ($\approx t/3$ in cuprates
\cite{HybertsenEtAl90}). Second-order superexchange is crucial for the physics
of the doped fermionic Mott-insulator. We want to emphasize that our proposal,
in which both spin coupling and hopping originate from superexchange, does not
involve any process above second order. Our scheme is, in a sense, contrary to
the slave-boson approach to the $t$-$J$ model (where auxiliary bosons
describing holes are introduced as a purely theoretical concept): We propose
a physical system of real bosons and spin 1/2 fermions that behaves as a system
of fermions alone.

Before presenting the details of our proposal, we want to mention that recently
another, interesting solution to problem (ii) of controlled hole doping 
[not to the issue of temperature, problem (i)] has been proposed \cite{HoEtAl09}:
The authors suggest realizing an attractive Hubbard model, which for a bipartite
lattice can be mapped to a repulsive one (see, e.g., \cite{Auerbach}). 
In particular, an imbalance of $\up$ and $\dw$ fermions for attractive
interaction, being relatively easy to control experimentally, corresponds to
hole or particle doping in the repulsive model. Unfortunately, at the same time,
the trap felt by the attractive fermions transforms to an inhomogeneous magnetic
field, favoring a spatial separation of repulsive $\up$ and $\dw$ fermions.

Let us now describe in detail our approach to controlled hole doping. We
consider a mixture of $N_f\equiv N_\up+N_\dw$ fermionic and $N_b$ bosonic atoms
($N\equiv N_f+N_b$) in an optical lattice, with the fermions having equally
populated ``spin'' states $s=\up,\dw$. The system is described quantitatively
by the Hubbard model~\cite{JakschEtAl98}  
\bes\label{eq:Hfb}
\Ho_{f\!fb} 
&=& -\sum_{\la i,j\ra}\Big[
		\sum_{s=\up,\dw}t^f_{ij} (\fa_{is}\fo_{js}+\text{h.c.})
	          + t^b_{ij}(\ba_i\bo_j + \text{h.c.})  \Big]
\nonumber\\ 	  
  &+&\sum_{i}  \big[  U^{f\!f}\no_{i\up}\no_{i,\dw} 
  		    + U^{fb}\no_{if}\no_{ib}
		    + \frac{U^{bb}}{2}\no_{ib}(\no_{ib}-1)\big] 
\nonumber\\ 	
  &+&\sum_{i}  \big[ V_i\no_i + \delta V_i \no_{if} \big].  
\ees
Here $\fa_{is}$ and $\ba_{i}$ denote creation operators for fermions and bosons at
site $i$. We also define number operators 
$\no_{is}\equiv\fa_{is}\fo_{is}$, $\no_{ib}\equiv\ba_i\bo_i$, 
$\no_{if}\equiv\no_{i\up}+\no_{i\dw}$, and $\no_i\equiv\no_{if}+\no_{ib}$. 
The first line of Eq.~(\ref{eq:Hfb}) comprises fermion and boson tunneling
between neighboring sites, with positive matrix elements $t^f_{ij}$ and $t^b_{ij}$
(depending exponentially on lattice depth \cite{JakschEtAl98}). The second line
takes care of the repulsive on-site interaction between the different types of
particles, with Hubbard energies $U^{f\!f}$, $U^{fb}$, and $U^{bb}$
(weakly depending on lattice depth and being proportional to the
corresponding $s$-wave scattering lengths \cite{JakschEtAl98}). Finally, the
third line includes co-centric trapping potentials $V^f_i\equiv V_i+\delta V_i$ and
$V^b_i\equiv V_i$ for fermions and bosons [$V^f_i\equiv V^b_i\equiv 0$ in the center], with 
$\delta V_i$ to be tuned small. 

We are interested in the parameter regime giving rise to an extended mixed Mott region, with 
one particle (boson or fermion) per site, in the center of the trap. We assume
a cubic lattice of spacing $d$ with site locations $d\br_i$, as well as a 
parabolic confinement $V_i=\frac{1}{2}\alpha \br_i^2$. We also define both 
$U_\text{min}\equiv \min(U^{f\!f},U^{fb},U^{bb})$ and 
$t_\text{max}=\max(\{t^f_{ij}\},\{t^b_{ij}\})$. Temperatures are well lower
than $t_\text{max}$. Now, $ t_\text{max} \ll U_\text{min}$ and 
$\mu<U_\text{min}$ (while $t_\text{max}\ll U_\text{min}-\mu$), with chemical
potential $\mu$, guarantees strong suppression of double occupancy. On the
other hand, the trap $V_i$ prevents vacancies from entering the central region:
the occupied region of radius $\rho\approx(3N/4\pi)^{\!1/3}$ features only a
thin  shell (a few $d$ wide) of reduced particle number, provided
$t_\text{max}\lesssim\alpha\rho\equiv \Delta V_\text{edge}$, where $\Delta 
V_\text{edge}$ is the energy difference between sites at 
radius $\rho$ and $\rho+1$. With the above
requirements fulfilled, the chemical potential for bosons and fermions is 
basically given by the potential energy needed to place a particle at the edge
of the occupied region, $\mu\approx \frac{1}{2}\alpha\rho^2$. One has 
$\Delta V_\text{edge}/\mu = 2\rho$. The radii $\rho$ might range from 10
to 30. All in all, 
$t_\text{max} \lesssim \Delta V_\text{edge}\ll \mu < U_\text{min}$ 
summarizes the parameter regime assumed here. In practice, 
$t_\text{max}/U_\text{min}$ can be adjusted via 
the lattice depths, while $\mu/U_\text{min}$ can be tuned to a
value of 1/2, say, by varying the trap depth $\alpha$.

%

In the bulk of the mixed Mott region, at each site $i$, strong repulsion in
combination with the trapping confinement $V_i$ gives rise to the constraint
	\be\label{eq:subspace}
	\no_i=\no_{i\up}+\no_{i\dw}+\no_{ib} = 1,
	\ee
with the overall boson fraction $x\approx N_b/N$. The system can be described by
an effective Hamiltonian $\Ho_\text{eff}$ acting in the subspace
$\mathcal{S}_1$ defined by Eq.~(\ref{eq:subspace}). Treating the first line
of $\Ho_{f\!fb}$ as perturbation $\Ho_1$, we can expand $\Ho_\text{eff}$
according to degenerate perturbation theory.\footnote{With unperturbed
states $a,a'\in\mathcal{S}_1$, $b\not\in\mathcal{S}_1$ and 
energies $E_a$, $E_{a'}$, $E_b$, the leading orders of $\Ho_\text{eff}$
read \cite{Klein73}:
$\la a'|\Ho^{(0)}_\text{eff}|a\ra =\la a'|\Ho_0|a\ra$,
$\la a'|\Ho^{(1)}_\text{eff}|a\ra =\la a'|\Ho_1|a\ra$, and
$\la a'|\Ho^{(2)}_\text{eff}|a\ra =\sum_b\la a'|\Ho_1|b\ra\la b|\Ho_1|a\ra
\frac{1}{2}[(E_a-E_b)^{-1}+(E_{a'}-E_b)^{-1}]$.}
Up to second order, using $\no_{ib}=1-\no_{if}$ in $\mathcal{S}_1$, 
one finds
	\bes\label{eq:sb}
	\Ho_\text{eff} 
	&=&\Po\Big\{ 
	-\sum_{\la ij\ra}\sum_s t_{ij}(\fa_{is}\bo_i\ba_j\fo_{js}+\text{h.c.})
	+ \sum_i W_i\no_{if}
	\nonumber\\
	&+&\sum_{\la ij\ra} \Big[J_{ij}\big(\So_i\So_j
		-\frac{\no_{if}\no_{jf}}{4}\big)
		+U^{nn}_{ij}\no_{if}\no_{jf} \Big]
		\Big\}\Po ,
	\ees
where $\Po$ projects on $\mathcal{S}_1$ and where we have introduced the usual spin operators $\So_i\equiv\frac{1}{2}\sum_{s's}\fa_{is'}{\bm\sigma}_{s's}\fo_{is}$ with Pauli matrices ${\bm\sigma}_{s's}$. Moreover,
	\bes\label{eq:t}
	t_{ij} &\equiv& 2 \frac{t^f_{ij}t^b_{ij}}{U^{fb}}	    
	\big[1+\delta^{fb}_{ij}\big],
	\\\label{eq:J}
	J_{ij} &\equiv& 4 \frac{(t^f_{ij})^2}{U^{f\!f}}	
	\big[1+\delta^{f\!f}_{ij}\big],
	\\\label{eq:Unn}
	U^{nn}_{ij}&\equiv& (I^{fb}_{ij}-I^{bb}_{ij}), 
	\\\label{eq:W}
	W_i &\equiv&  \delta V_i +\sum_{j\in\text{adj}(i)} \big[I^{bb}_{ij}
			-I^{fb}_{ij}/2-\delta I^{fb}_{ij}\big].
	\ees
with $\delta^{\nu}_{ij}\equiv [(\frac{U^{\nu}}{V_i-V_j})^2-1]^{\!-1}$,
$I^{bb}_{ij}\equiv 4\frac{(t^b_{ij})^2}{U^{bb}}[1+\delta^{bb}_{ij}]$, 
$I^{fb}_{ij} \equiv 2\frac{(t^f_{ij})^2+(t^b_{ij})^2}{U^{fb}}[1+\delta^{fb}_{ij}]$,
$\delta I^{fb}_{ij} \equiv \frac{(t^f_{ij})^2-(t^b_{ij})^2}{U^{fb}} 
\delta^{fb}_{ij}\frac{U^{fb}}{V_i-V_j}$, and $\text{adj}(i)$ containing all 
sites adjacent to $i$.
In Hamiltonian (\ref{eq:sb}), $t_{ij}$ describes boson-fermion swaps, i.e.\ 
effective fermion tunneling,
$W_i$ is the effective potential felt by fermions, $J_{ij}$ stands for the usual
fermionic antiferromagnetic superexchange coupling, and $U_{ij}^\text{nn}$
captures boson-mediated nearest-neighbor interaction between
fermions that can be attractive, repulsive or zero.

Hamiltonian (\ref{eq:sb}) is equivalent to 
a $t$-$J$-type model, describing a purely fermionic system when strong repulsion
suppresses double occupancy. This can be seen by introducing 
composite-fermion creation operators
	\be\label{eq:trans}
	\ca_{is}\equiv\fa_{is}\bo_i.
	\ee
We define $\tilde{n}_{is}\equiv\ca_{is}\co_{is}$ and
$\tilde{\bm S}_i\equiv\frac{1}{2}\sum_{s's}\ca_{is'}{\bm\sigma}_{s's}\co_{is}$. Using
$\Po\fa_{is'}\fo_{is}\Po= \Po\fa_{is'}(1+\ba_i\bo_i)\fo_{is}\Po=
\Po\ca_{is'}\co_{is}\Po$, yields both
$\tilde{n}_{is}=\no_{is}$ and $\tilde{\bm S}_i=\So_i$ in $\mathcal{S}_1$.
Bosons transform to empty sites (holes) and $\mathcal{S}_1$ to the subspace
defined by $\tilde{n}_i\le1$ with 
$\tilde{n}_{i}\equiv\tilde{n}_{i\up}+\tilde{n}_{i\dw}$.
Thus, we can rewrite $\Ho_\text{eff}$ as $t$-$J$ Hamiltonian in terms of
$\ca_{is}$-fermions alone:
	\bes\label{eq:tJ}
	\Ho_\text{eff}
	&=&\Po\Big\{
	     -\sum_{\la ij\ra}\sum_s t_{ij}(\ca_{is}\co_{js}+\text{h.c.})
		+\sum_iW_i\tilde{n}_{i}
	\nonumber\\
	   &+&\sum_{\la ij\ra} J_{ij}\Big[\big(\tilde{\bm S}_i\tilde{\bm S}_j
			 -\frac{\tilde{n}_{i}\tilde{n}_{j}}{4}\big)
			 +U^{nn}_{ij}\tilde{n}_{i}\tilde{n}_{j}\Big]
	    \Big\}\Po.
	\ees
Transformation (\ref{eq:trans}) (being inverse to a the slave-boson transformation)
is illustrated in Fig.~\ref{fig:principle}. 
	
The $t$-$J$ model (\ref{eq:tJ}) with $W_i=\text{const.}$, $U^{nn}_{ij}=0$,
$t_{ij}=t$, $J_{ij}=J\approx t/3$, and boson fraction $x$, is the most simple
candidate to explain high-temperature cuprate superconductivity 
\cite{Anderson87,AndersonEtAl04, LeeEtAl06,HybertsenEtAl90}. The fermion-boson
mixture considered here, allows one to realize these parameters quite accurately:
Starting from a cubic lattice, the tunneling amplitudes $t^f_{ij}$ and
$t^b_{ij}$ can be suppressed in one direction by ramping up the lattice in that
direction, leading to a stack of uncoupled square lattices layers. 
From now on, we will only consider the intra-layer physics.
Creating the optical lattice by using a rather broad laser beam, within the occupied region 
of the trap (created by further beams) the lattice will be practically homogeneous.  Thus, 
$t^f_{ij}\simeq t^f$ and $t^b_{ij}\simeq t^b$. Furthermore, potential differences
$|V_i-V_j|$ between neighboring sites are much smaller than
$U^{f\!f}$, $U^{fb}$ and $U^{bb}$, giving $\delta_{ij}^\nu\lesssim 10^{-3}$
($\nu={f\!f},{fb},{bb}$) for the parameters estimated in a previous
paragraph. With that the model parameters (\ref{eq:t})-(\ref{eq:Unn}) are to good
approximation homogeneous within the simulator region, $t_{ij}\simeq t$, $J_{ij}\simeq J$,
and $U^{nn}_{ij}\simeq U^{nn}$. 
By the very same arguments, also the last three terms contributing to $W_i$ 
[Eq.~(\ref{eq:W})] have a negligible spatial dependence compared to the effective
hopping parameter $t$, being the relevant energy scale here. Thus, if also the
difference between boson and fermion potential $\delta V_i$ is smaller than
$t$, one achieves a practically flat effective potential $W_i\simeq W$,
without fermions leaking out of the system, cf.\ Fig.~\ref{fig:principle}(b).

In an experimental realization, the hole fraction~$x$ is controlled by the
boson fraction (also amenable to postselection). Moreover,
$t/J \simeq \tau u_f/2$ and $U^\text{nn}/J\simeq [2u_b^{-1}-\tau^{-2}-1]u_f/2$,
with $\tau\equiv t^b/t^f$ and $u_{\nu}\equiv U^{\nu\nu}/U^{fb}$ ($\nu=f,b$),
can be tuned by using Feshbach resonances and by modifying the lattices for
bosons and fermions, either in depth or relative position.
A candidate system is a mixture of Ytterbium isotopes
\cite{YbFermions,YbSL,YbFeshbach}: The total angular momentum of Yb is just 
given by the nuclear spin $I$, not influencing the interparticle interaction. 
A mixture of two spin states of fermionic $^{173}$Yb ($I=5/2$) with bosonic 
$^{168}$Yb or $^{174}$Yb (both $I=0$) is described by three positive $s$-wave
scattering lengths, with $a_{f\!f}/a_{fb}=5.2$ (1.4), $a_{bb}/a_{fb}=6.5$
(0.8) and $a_{fb}=$ 2.0nm (7.3nm) for $^{168}$Yb ($^{174}$Yb) \cite{YbSL}.
Also optical Feshbach resonances are available \cite{YbFeshbach}. Using a far
off-resonant optical potential allows one to create practically equal traps for
bosonic and fermionic isotopes without fine-tuning. Considering the Mott 
regime, according to band structure calculations, typical Hubbard interaction 
parameters for Yb will roughly be on the order of 10 kHz. The effective 
tunneling matrix element $t$ will be about 200 times smaller. 
Such low energy scales are challenging experimentally. However, they are not
very specific to our proposal, but rather are a generic consequence of the fact 
that the $t$-$J$ physics is inevitably the physics of superexchange processes.
Therefore, the temperatures needed here are only moderately lower than those
that would be needed in an experimental setup where---somehow---a Mott insulator
of fermionic atoms can be doped with real holes in a controlled fashion. While
in the latter case $U^{f\!f}/t^f\gg1$ would be required, in our proposal one has
$U^{f\!f}/t^f\gg 2(t/J)\approx 6$ (combine $\tau u_f\simeq 2t/J$ and 
$U^{fb}\gg t^b$), reducing $J$ by a factor of 6 only.

At hole doping $x\lesssim\mathcal{O}(0.01)$ and temperatures
$k_\text{B}T\lesssim J$, the homogeneous $t$-$J$ model (\ref{eq:tJ}) 
is known to give rise to (quasi-)long-range antiferromagnetic N\'eel order. 
At larger doping $x\sim\mathcal{O}(0.1)$ and even lower temperatures, the model
is conjectured to possess yet another ordered phase, namely the
$d_{x^2-y^2}$-wave superconducting one observed also in the cuprates
\cite{AndersonEtAl04,LeeEtAl06}. Superfluidity of $\hat{c}$-fermions 
(that is in the $t$-$J$ model) is connected to superfluidity of \emph{both}
$\hat{f}$-fermions and $\hat{b}$-bosons \cite{IoffeLarkin89}; both have to be
probed. The bosonic superfluidity is related to a (quasi)condensate
at quasimomentum $\bk=0$, visible in time-of-flight absorption images.\footnote{
Bosons and $\up$ and $\dw$ fermions can be imaged separately by
selective absorption or Stern-Gerlach separation.}. In order to verify $d$-wave
superfluidity of the fermions one can resort to different schemes for measuring
quasiparticle excitations (as well as their interference) that have been
proposed for cold atom systems for this purpose \cite{HofstetterEtAl02,Measure}.

In the cuprates, at optimal doping, superconducting behavior appears at
temperatures that are at least one order of magnitude smaller than
$J$ \cite{LeeEtAl06,HybertsenEtAl90}. It is challenging to achieve these
temperatures with cold atoms, and novel cooling techniques, as those proposed in
Refs.~\cite{AdvancedCooling}, have to be implemented. 
The system has to be divided into a low-entropy part (lacking low-lying
excitations) and a part carrying most of the entropy (a metallic shell or a
boson-gas reservoir). The latter is removed before the system is adiabatically
transferred into the desired state. In our fermion-boson system, a suitable gapped
low-entropy phase would be an insulator of two fermions and an integer number of
bosons at each site, to be created by making the trap steep and the lattice
deep.

Finally, we would like to mention that neutralizing the trapping potential the
way described here is also possible and interesting for bosons. In Refs.\
\cite{Trap} it has been shown that the presence of a trap can fundamentally
change the nature of the transition between an insulating and a superfluid
quantum phase. The reason is not that the trapped system is too small to
sharply distinguish between the two phases. It is more subtle: Because of the
inhomogeneity, the transition occurs locally by displacing the interface
separating spatial regions of different bulk phases. However,
this spatial interface might be a smooth crossover rather than
a sharp transition. A direct consequence can be that the adiabatic passage
between two quantum phases is greatly facilitated by the trap. A simple
lattice model that would allow to study such phenomena is given by spinless
hard-core bosons on a square lattice with nearest-neighbor repulsion
$U_{nn}$ and hopping $t$. It features a transition from a checkerboard
insulator at half filling and small $t/U_{nn}$ to a superfluid phase
\cite{HebertEtAl01}. The model (being equivalent to an XXZ spin 1/2 model)
can be realized experimentally by creating a mixed Mott insulator having
one boson of either species $s= 1$ or 2 on each site \cite{KuklovSvistunov03}. 
Mapping $\ba_i\equiv\ba_{i1}\bo_{i2}$, the system
is described by a single type of hard-core boson only, with 
$U_{nn}=(t_1^2+t_2^2)/U_{12}-4t_2^2/U_{22}$, $t=2t_1t_2/U_{12}$ and residual 
trapping potential $W_i=V_{i1}-V_{i2}$. Here $U_{ss'}$, $t_s$, and $V_{is}$ denote
on-site interaction, tunneling, and traps for the species $s=1$ and 2.
Now $W_i$ can be tuned constant without making particles 
leak out. Thus, a ramp of $t/U_\text{nn}$ at different rates can now be performed 
both with and without parabolic potential, and the degree of adiabaticity in
passing the transition can be compared.

We have proposed a robust implementation of a quantum simulator
for the homogeneous $t$-$J$ model with well controlled hole doping, using a
sample of ultracold bosonic and fermionic atoms in an optical lattice. We
believe that---once the necessary temperatures are realized---our scheme can
serve to gain crucial insight into the physics of strongly correlated quantum
matter. Moreover, realizing a \emph{homogeneous} bosonic system allows one to
investigate experimentally the role of inhomogeneity when passing from one quantum
phase to another.

We thank P.\ Massignan, A.\ Muramatsu, V.\ Ahufinger, and A.\
Sanpera for discussion.
Support by the Spanish MICINN (FIS2008-00784, FIS2007-29996-E), the A.\ v.\
Humboldt Foundation, ERC Grant QUAGATUA, and EU STREP NAMEQUAM is acknowledged.


\begin{thebibliography}{platz}
\bibitem{Lattices}
M.\ Lewenstein, A.\ Sanpera, V.\ Ahufinger, B.\ Damski, A.\ Sen(de), and U.\ Sen,
Adv.\ Phys.\ {\bf 56}, 243 (2007);
I.\ Bloch, J.\ Dalibard, and W.\ Zwerger, Rev.\ Mod.\ Phys.\ {\bf 80}, 885 (2008). 
\bibitem{JakschEtAl98}
D.\ Jaksch, C.\ Bruder, J.I.\ Cirac, C.W.\ Gardiner, and P.\ Zoller,
Phys. Rev. Lett. {\bf 81}, 3108 (1998).
\bibitem{Feynman82}
R.\ Feynman, Int.\ J.\ Theor.\ Phys.\ {\bf21},467 (1982).
\bibitem{FisherEtAl89}
M.P.A.\ Fisher, P.B.\ Weichman, G.\ Grinstein, and D.S.\ Fisher,
Phys.\ Rev.\ B {\bf 40}, 546 (1989). 
\bibitem{GreinerEtAl02}
M.\ Greiner, O.\ Mandel, T.\ Esslinger, T.W.\ H\"ansch, and I.\ Bloch,
Nature {\bf 415}, 39 (2002). 
\bibitem{TrotzkyEtAl09}
S.\ Trotzky, L.\ Pollet, F.\ Gerbier, U.\ Schnorrberger, I.\ Bloch,
		N.\ Prokof'ev, B.\ Svistunov, and M.\ Troyer, 
arXiv:0905.4882 (2009).
\bibitem{Antiferro} 
C. Lhuillier, arXiv:cond-mat/0502464 (2005);
F. Alet, A.M. Walczak, and M.P.A.\ Fisher, Physica A {\bf 369}, 122 (2006);
R.\ Moessner and A.P. Ramirez, Physics Today {\bf 59-2}, 24 (2006);
S. Sachdev, Nat.\ Phys.\ {\bf 4}, 173 (2008).
\bibitem{Anderson87}
P.W.\ Anderson, Science {\bf235}, 1196 (1987).
\bibitem{AndersonEtAl04}
P.W.\ Anderson, P.A.\ Lee, M.\ Randeria, T.M.\ Rice, N.\ Trivedi,
and F.C.\ Zhang, J.\ Phys.: Condens.\ Matter {\bf16}, R755 (2004);
K.\ Le Hur and T.M.\ Rice, Ann.\ Phys.\ (New York) {\bf 324}, 1452 (2009).
\bibitem{LeeEtAl06}
P.A.\ Lee, N.\ Nagaosa, and X.-G.\ Wen, Rev.\ Mod.\ Phys.\ {\bf78}, 17 (2006).
\bibitem{HofstetterEtAl02}
W.\ Hofstetter, J.I.\ Cirac, P.\ Zoller, E.\ Demler, and M.D.\ Lukin,
Phys.\ Rev.\ Lett.\ {\bf 89}, 220407 (2002).
\bibitem{Mott}
R.\ J\"ordens, N.\ Strohmaier, K.\ G\"unter, H.\ Moritz, and T.\ Esslinger,
Nature {\bf 455}, 204 (2008);
U.\ Schneider, L.\ Hackerm\"uller, S.\ Will, T.\ Best, I.\ Bloch, T.A.\ Costi,
R.W.\ Helmes, D.\ Rasch, and A.\ Rosch, 
Science {\bf322}, 1520 (2008).
\bibitem{AdiabaticCooling}
F.\ Werner, O.\ Parcollet, A.\ Georges, and S.R.\ Hassan,
Phys.\ Rev.\ Lett.\ {\bf95}, 056401 (2005);
A.-M.\ Dar\'e, L.\ Raymond, G.\ Albinet, and A.-M.S.\ Tremblay, 
Phys.\ Rev.\ B {\bf76}, 064402 (2007);
A.\ Koetsier, R.A.\ Duine, I.\ Bloch, and H.T.C.\ Stoof,
Phys.\ Rev.\ A {\bf77}, 023623 (2008). 
\bibitem{AdvancedCooling}
J.\ Bernier, C.\ Kollath, A.\ Georges, L.D.\ Leo, F.\ Gerbier, 
	C.\ Salomon, and M.\ K\"ohl, 
Phys.\ Rev.\ A {\bf79}, 061601(R) (2009);
T.-L.\ Ho and Q.\ Zhou, Proc.\ Natl.\ Acad.\ Sci.\ USA {\bf106}, 6916 (2009).
\bibitem{EckardtEtAl09b}
A cold atom realization of an (undoped) quantum antiferromagnet, with the
coupling large compared to presently available temperatures, has been
proposed in: A.\ Eckardt, P.\ Hauke, P.\ Soltan-Panahi, C.\ Becker,
K.\ Sengstock, and M.\ Lewenstein, EPL {\bf89}, 10010 (2010). 
\bibitem{Trap}
G.G.\ Batrouni, V.\ Rousseau, R.T.\ Scalettar, M. Rigol, A. Muramatsu,
P.J.H.\ Denteneer, M.\ Troyer, Phys.\ Rev.\ Lett.\ {\bf89}, 117203 (2002);
S.\ Wessel, F.\ Alet, M.\ Troyer, and G.G.\ Batrouni Phys.\ Rev.\ A {\bf70},
 053615 (2004).
\bibitem{KuklovSvistunov03} 
A.B.\ Kuklov and B.V.\ Svistunov, Phys.\ Rev.\ Lett.\ {\bf90}, 100401 (2003);
\bibitem{LewensteinEtAl04} 
M. Lewenstein et al., Phys.\ Rev.\ Lett.\  {\bf 92}, 050401 (2004).
\bibitem{ZouAnderson88} J. Spa{\l}ek and A. M. Ole\'s, Physica B {\bf 86-88}, 375 (1977);
Z.\ Zou and P.W.\ Anderson, Phys.\ Rev.\ B {\bf37}, 627 (1988).
\bibitem{Duan05Anna}
L.-M. Duan, Phys.\ Rev.\ Lett.\ {\bf 95}, 243202 (2005);
 A.\ Muramatsu, V.\ Ahufinger, and A. Sanpera, in preparation. 
\bibitem{HybertsenEtAl90}
M.S.\ Hybertsen, E.B.\ Stechel, M.\ Schluter, and D.R.\ Jennison,
Phys.\ Rev.\ B {\bf41}, 11068 (1990).
\bibitem{HoEtAl09}
A.F.\ Ho, and M.A.\ Cazalilla, and T.\ Giamarchi, Phys.\ Rev.\ A {\bf 79},
033620 (2009).
\bibitem{Auerbach} A.\ Auerbach, \emph{Interacting Electrons and Quantum
Magnetism} (Springer-Verlag, New York, 1994).
\bibitem{Klein73}
D.J.\ Klein, J.\ Chem.\ Phys.\ {\bf61}, 786 (1973). 
\bibitem{YbFermions}
T.\ Fukuhara, Y.\ Takasu, M.\ Kumakura, and Y.\ Takahashi,
Phys.\ Rev.\ Lett.\ {\bf 98}, 030401 (2007);  
\bibitem{YbSL}
M.\ Kitagawa et al.\, Phys.\ Rev.\ A {\bf 77}, 012719 (2008); 
\bibitem{YbFeshbach}
K.\ Enomoto, K.\ Kasa, M.\ Kitagawa, and Y.\ Takahashi,
Phys.\ Rev.\ Lett.\ {\bf 101}, 203201 (2008). 
\bibitem{IoffeLarkin89}
L.B.\ Ioffe and A.I.\ Larkin,
Phys.\ Rev.\ B {\bf39}, 8988 (1989). 
\bibitem{Measure}
T.-L.\ Dao, A.\ Georges, J.\ Dalibard, C.\ Salomon, and I.\ Carusotto,
Phys.\ Rev.\ Lett.\ {\bf98}, 240402 (2007);
D.\ Pekker, R.\ Sensarma, and E.\ Demler, arXiv:0906.0931.  
\bibitem{HebertEtAl01}
F.\ H\'ebert, G.G.\ Batrouni, R.T.\ Scalettar, G.\ Schmid, M.\ Troyer,
A.\ Dorneich, Phys.\ Rev.\ B {\bf65}, 014513 (2001).
\end{thebibliography}
\end{document}